\documentclass[aps, prl, twocolumn, showpacs, amsmath, amssymb, superscriptaddress]{revtex4}
\usepackage{graphicx}
\usepackage{dcolumn}
\usepackage{bm}
\usepackage{alltt}
\usepackage{calc,rotating,xspace}
\usepackage{pifont}
\usepackage{axodraw}
\usepackage{hhline}
\newcommand{\ppbar} {\ensuremath{p \overline p}}
\newcommand{\Bs} {B_s^0}
\newcommand{\aB} {B_{(s)}^0}
\newcommand{\aD} {D_{(s)}^-}
\newcommand{\Ds} {D_s^-}
\newcommand{\Bd} {B^0}
\newcommand{\Dd}  {D^-}

\newcommand{\Kp} {K^+}

\newcommand{\tripi}{\pi^{+}\pi^{+}\pi^{-}}
\newcommand{\pip}{\pi^{+}} 
\newcommand{\pim}{\pi^{-}} 
\newcommand{\pipm}{\pi^{+}\pi^{-}}

\newcommand{\keV}{\ensuremath{\rm ke\kern -0.1em V}}

\newcommand{\MeV}{\ensuremath{\rm Me\kern -0.1em V}}
\newcommand{\MeVc}{\ensuremath{\MeV\!/c}}

\newcommand{\GeV}{\ensuremath{\rm Ge\kern -0.1em V}}
\newcommand{\TeV}{\ensuremath{\rm Te\kern -0.1em V}}
\newcommand{\GeVc}{\ensuremath{\GeV\!/c}}

\newcommand{\stat}{(\mathrm{stat.})}
\newcommand{\syst}{(\mathrm{syst.})}
\newcommand{\br}{(\mathrm{br})}
\newcommand{\pr}{(\mathrm{pr})}

\newcommand{\mm}{\ensuremath{\rm mm}}

\newcommand{\ipb}{\ensuremath{\mathrm{pb^{-1}}}}

\newcommand{\nBsDsPiTotal}{494}
\newcommand{\eBsDsPiTotal}{28}

\newcommand{\nBdDPiTotal}{8098}
\newcommand{\eBdDPiTotal}{114}

\newcommand{\nBsDsTriPiTotal}{160}
\newcommand{\eBsDsTriPiTotal}{17}
\newcommand{\nBdDTriPiTotal}{3288}
\newcommand{\eBdDTriPiTotal}{76}

\newcommand{\nBsKSKTotal}{90}
\newcommand{\eBsKSKTotal}{17}

\newcommand{\nBsTriPiTotal}{49}
\newcommand{\eBsTriPiTotal}{11}

\newcommand{\fsOverfd}{0.259}
\newcommand{\fsOverfdError}{0.038}

\newcommand{\bsdspiResCorr}{1.13}
\newcommand{\bsdspiStatCorr}{0.08}
\newcommand{\bsdspiSystCorr} {0.05}
\newcommand{\bsdspiEBrCorr}{0.15}
\newcommand{\bsdspiEPrCorr}{0.17}
\newcommand{\bsdspiSystTotCorr}{0.23}
\newcommand{\Colangelo}{1.05}
\newcommand{\ColangeloErr}{0.24}

\newcommand{\bsdstripiComboResCorr}{1.05}
\newcommand{\bsdstripiComboStatCorr}{0.10}
\newcommand{\bsdstripiComboSystCorr}{0.07}
\newcommand{\bsdstripiComboBrCorr}{0.14}
\newcommand{\bsdstripiComboPrCorr}{0.15}
\newcommand{\bsdstripiComboSystTotCorr}{0.22}
\newcommand{\effRatiodspi}{0.913}
\newcommand{\effRatiodspiErr}{0.004}
\newcommand{\effRatiodstripi}{0.814}
\newcommand{\effRatiodstripiErr}{0.010}
\newcommand{\effRatiodsksk}{0.352}
\newcommand{\effRatiodskskErr}{0.009}
\newcommand{\effRatiodsTripi}{0.397}
\newcommand{\effRatiodsTripiErr}{0.009}

\newcommand{\dpdsphipir}{4.40} 
\newcommand{\dpdsphipirErr}{0.59} 
\newcommand{\dpdskskr}{3.80} 
\newcommand{\dpdskskrErr}{0.77} 
\newcommand{\dpdstripir}{7.80} 
\newcommand{\dpdstripirErr}{1.50} 

\newcommand{\bsdspiRes}{0.292}
\newcommand{\bsdspiResStat}{0.020}
\newcommand{\bsdspiResSyst}{0.012}
\newcommand{\bsdstripiRes}{0.263}
\newcommand{\bsdstripiResStat}{0.029}
\newcommand{\bsdstripiResSyst}{0.018}
\newcommand{\bsdskskRes}{0.274}
\newcommand{\bsdskskResStat}{0.053}
\newcommand{\bsdskskResSyst}{0.030}
\newcommand{\bsdsTripiRes}{0.293}
\newcommand{\bsdsTripiResStat}{0.067}
\newcommand{\bsdsTripiResSyst}{0.021}
\newcommand{\brBsDsPi}{3.8}
\newcommand{\brBsDsPiStatErr}{0.3}
\newcommand{\brBsDsPiSystErr}{1.3}
\newcommand{\brBsDsTriPi}{8.4}
\newcommand{\brBsDsTriPiStatErr}{0.8}
\newcommand{\brBsDsTriPiSystErr}{3.2}
\begin{document}
\title{
\boldmath
 Measurement of the Ratios of Branching Fractions 
$\mathcal{B}(\Bs \to \Ds \tripi) / \mathcal{B}(\Bd \to \Dd \tripi)$ and 
$\mathcal{B}(\Bs \to \Ds \pip) / \mathcal{B}(\Bd \to \Dd \pip)$
}
\pacs{13.25.Hw 14.40Nd}
\affiliation{Institute of Physics, Academia Sinica, Taipei, Taiwan 11529, Republic of China} 
\affiliation{Argonne National Laboratory, Argonne, Illinois 60439} 
\affiliation{Institut de Fisica d'Altes Energies, Universitat Autonoma de Barcelona, E-08193, Bellaterra (Barcelona), Spain} 
\affiliation{Baylor University, Waco, Texas  76798} 
\affiliation{Istituto Nazionale di Fisica Nucleare, University of Bologna, I-40127 Bologna, Italy} 
\affiliation{Brandeis University, Waltham, Massachusetts 02254} 
\affiliation{University of California, Davis, Davis, California  95616} 
\affiliation{University of California, Los Angeles, Los Angeles, California  90024} 
\affiliation{University of California, San Diego, La Jolla, California  92093} 
\affiliation{University of California, Santa Barbara, Santa Barbara, California 93106} 
\affiliation{Instituto de Fisica de Cantabria, CSIC-University of Cantabria, 39005 Santander, Spain} 
\affiliation{Carnegie Mellon University, Pittsburgh, PA  15213} 
\affiliation{Enrico Fermi Institute, University of Chicago, Chicago, Illinois 60637} 
\affiliation{Joint Institute for Nuclear Research, RU-141980 Dubna, Russia} 
\affiliation{Duke University, Durham, North Carolina  27708} 
\affiliation{Fermi National Accelerator Laboratory, Batavia, Illinois 60510} 
\affiliation{University of Florida, Gainesville, Florida  32611} 
\affiliation{Laboratori Nazionali di Frascati, Istituto Nazionale di Fisica Nucleare, I-00044 Frascati, Italy} 
\affiliation{University of Geneva, CH-1211 Geneva 4, Switzerland} 
\affiliation{Glasgow University, Glasgow G12 8QQ, United Kingdom} 
\affiliation{Harvard University, Cambridge, Massachusetts 02138} 
\affiliation{Division of High Energy Physics, Department of Physics, University of Helsinki and Helsinki Institute of Physics, FIN-00014, Helsinki, Finland} 
\affiliation{University of Illinois, Urbana, Illinois 61801} 
\affiliation{The Johns Hopkins University, Baltimore, Maryland 21218} 
\affiliation{Institut f\"{u}r Experimentelle Kernphysik, Universit\"{a}t Karlsruhe, 76128 Karlsruhe, Germany} 
\affiliation{High Energy Accelerator Research Organization (KEK), Tsukuba, Ibaraki 305, Japan} 
\affiliation{Center for High Energy Physics: Kyungpook National University, Taegu 702-701, Korea; Seoul National University, Seoul 151-742, Korea; and SungKyunKwan University, Suwon 440-746, Korea} 
\affiliation{Ernest Orlando Lawrence Berkeley National Laboratory, Berkeley, California 94720} 
\affiliation{University of Liverpool, Liverpool L69 7ZE, United Kingdom} 
\affiliation{University College London, London WC1E 6BT, United Kingdom} 
\affiliation{Centro de Investigaciones Energeticas Medioambientales y Tecnologicas, E-28040 Madrid, Spain} 
\affiliation{Massachusetts Institute of Technology, Cambridge, Massachusetts  02139} 
\affiliation{Institute of Particle Physics: McGill University, Montr\'{e}al, Canada H3A~2T8; and University of Toronto, Toronto, Canada M5S~1A7} 
\affiliation{University of Michigan, Ann Arbor, Michigan 48109} 
\affiliation{Michigan State University, East Lansing, Michigan  48824} 
\affiliation{Institution for Theoretical and Experimental Physics, ITEP, Moscow 117259, Russia} 
\affiliation{University of New Mexico, Albuquerque, New Mexico 87131} 
\affiliation{Northwestern University, Evanston, Illinois  60208} 
\affiliation{The Ohio State University, Columbus, Ohio  43210} 
\affiliation{Okayama University, Okayama 700-8530, Japan} 
\affiliation{Osaka City University, Osaka 588, Japan} 
\affiliation{University of Oxford, Oxford OX1 3RH, United Kingdom} 
\affiliation{University of Padova, Istituto Nazionale di Fisica Nucleare, Sezione di Padova-Trento, I-35131 Padova, Italy} 
\affiliation{LPNHE, Universite Pierre et Marie Curie/IN2P3-CNRS, UMR7585, Paris, F-75252 France} 
\affiliation{University of Pennsylvania, Philadelphia, Pennsylvania 19104} 
\affiliation{Istituto Nazionale di Fisica Nucleare Pisa, Universities of Pisa, Siena and Scuola Normale Superiore, I-56127 Pisa, Italy} 
\affiliation{University of Pittsburgh, Pittsburgh, Pennsylvania 15260} 
\affiliation{Purdue University, West Lafayette, Indiana 47907} 
\affiliation{University of Rochester, Rochester, New York 14627} 
\affiliation{The Rockefeller University, New York, New York 10021} 
\affiliation{Istituto Nazionale di Fisica Nucleare, Sezione di Roma 1, University of Rome ``La Sapienza," I-00185 Roma, Italy} 
\affiliation{Rutgers University, Piscataway, New Jersey 08855} 
\affiliation{Texas A\&M University, College Station, Texas 77843} 
\affiliation{Istituto Nazionale di Fisica Nucleare, University of Trieste/\ Udine, Italy} 
\affiliation{University of Tsukuba, Tsukuba, Ibaraki 305, Japan} 
\affiliation{Tufts University, Medford, Massachusetts 02155} 
\affiliation{Waseda University, Tokyo 169, Japan} 
\affiliation{Wayne State University, Detroit, Michigan  48201} 
\affiliation{University of Wisconsin, Madison, Wisconsin 53706} 
\affiliation{Yale University, New Haven, Connecticut 06520} 
\author{A.~Abulencia}
\affiliation{University of Illinois, Urbana, Illinois 61801}
\author{J.~Adelman}
\affiliation{Enrico Fermi Institute, University of Chicago, Chicago, Illinois 60637}
\author{T.~Affolder}
\affiliation{University of California, Santa Barbara, Santa Barbara, California 93106}
\author{T.~Akimoto}
\affiliation{University of Tsukuba, Tsukuba, Ibaraki 305, Japan}
\author{M.G.~Albrow}
\affiliation{Fermi National Accelerator Laboratory, Batavia, Illinois 60510}
\author{D.~Ambrose}
\affiliation{Fermi National Accelerator Laboratory, Batavia, Illinois 60510}
\author{S.~Amerio}
\affiliation{University of Padova, Istituto Nazionale di Fisica Nucleare, Sezione di Padova-Trento, I-35131 Padova, Italy}
\author{D.~Amidei}
\affiliation{University of Michigan, Ann Arbor, Michigan 48109}
\author{A.~Anastassov}
\affiliation{Rutgers University, Piscataway, New Jersey 08855}
\author{K.~Anikeev}
\affiliation{Fermi National Accelerator Laboratory, Batavia, Illinois 60510}
% ...
\collaboration{CDF Collaboration}
\noaffiliation

\begin{abstract}
Using 355 $\ipb$ of data collected by the CDF II detector in $\ppbar$ 
collisions at 
$\sqrt{s} = 1.96~\TeV$ at the Fermilab Tevatron, we study the fully 
reconstructed hadronic decays $\aB \to \aD \pip$ and  $\aB \to \aD \tripi$.  
We present the first measurement of the ratio of branching fractions 
$\mathcal{B}(\Bs \to \Ds \tripi) / \mathcal{B}(\Bd \to \Dd \tripi)$ = 
$\bsdstripiComboResCorr \pm \bsdstripiComboStatCorr \stat 
\pm \bsdstripiComboSystTotCorr \syst$. We also update our measurement of 
$\mathcal{B}(\Bs \to \Ds \pip) / \mathcal{B}(\Bd \to \Dd \pip)$ to 
$ \bsdspiResCorr \pm \bsdspiStatCorr \stat \pm \bsdspiSystTotCorr \syst$ 
improving the statistical uncertainty by more than a factor of two. 
We find  $\mathcal{B}(\Bs \to \Ds \pip) = [\brBsDsPi \pm \brBsDsPiStatErr 
\stat \pm \brBsDsPiSystErr \syst] \times 10^{-3}$ and 
$\mathcal{B}(\Bs \to \Ds \tripi) = [\brBsDsTriPi \pm 
\brBsDsTriPiStatErr \stat \pm \brBsDsTriPiSystErr \syst] \times 10^{-3}$.   
\end{abstract}

\maketitle
	Hadronic $B$ meson decays provide important information on both 
weak and hadronic interactions of heavy flavored mesons. The dominant 
hadronic decay modes of the $B$ meson involve tree-level diagrams where the 
$b \to c$ transition leads to a charmed meson and a virtual $W$ boson, 
which often emerges as a charged $\pi, \rho $, or $a_{1}(1260)$ 
meson \cite{LIPKIN}. The measurement of the ratios of branching fractions 
$\mathcal{B}(\Bs \to \Ds \pip [\pipm])$/ 
$\mathcal{B}(\Bd \to \Dd \pip [\pipm])$ \cite{CONJ} reveals information 
about $B$ decay mechanisms. One can attempt to separate the contributions 
of various processes in $\Bd \to \Dd \pip$ decay and then predict 
$\Bs \to \Ds \pip$ branching fraction using SU(3)\cite{DIAGRAMS, STEWART}, 
and further estimate flavor SU(3) symmetry breaking effects \cite{SU3}, 
which can be sizable \cite{FACTORIZATION}.

Experimentally the sample of $\Bs \to \Ds \tripi$ decays is included
in the data set used for the first measurement of the $\Bs$ oscillation 
frequency \cite{KROLL}. In this letter we present the first measurement 
of the ratio of branching fractions
$R(D 3\pi) = \mathcal{B}(\Bs \to \Ds \tripi) / \mathcal{B}(\Bd \to \Dd \tripi)$ using $\Ds$ decays to $\phi \pim$, $K^{*0} K^{-}$, $\pim \pim \pip$, and  
$\Dd$ decays to $\Kp \pim \pim$. We also update our previous \cite{IVAN} 
measurement of the ratio $R(D \pi) = \mathcal{B}(\Bs \to \Ds \pip)$/$\mathcal{B}(\Bd \to \Dd \pip)$. We measure the ratios of branching fractions because 
most of the systematic uncertainties cancel due to the similarity 
of final state kinematics. In addition, some common theoretical 
factors \cite{LEIBOVICH} also cancel in taking the ratio.  
The measurement is performed  using a sample of inclusive heavy flavor decays
selected by the displaced track trigger described below, corresponding to 
an integrated luminosity of  355 $\ipb$ of $\ppbar$ collisions 
at $\sqrt{s} = 1.96$ TeV. 

The components of the CDF II detector relevant for this analysis are 
briefly described below. A more complete description can be found 
elsewhere \cite{CDF}. Charged particle tracks are reconstructed in 
the pseudo-rapidity range $|\eta| \lesssim 1.0$, where $\eta$ is 
defined as $ - \ln \tan(\theta/2)$, and $\theta$ represents the angle 
between the particle and the proton beam direction \cite{REFSYSTEM}. 
Tracks are reconstructed from hits in the silicon microstrip detector (SVX) 
and the central outer tracker (COT). Both detectors are inside a 1.4 T 
solenoidal magnetic field. The SVX detector is composed of L00 
(single layer of silicon placed close to the beam pipe), SVX II 
(five cylindrical layers of double-sided sensors), and ISL (outermost layer 
of silicon), providing up to 8 coordinate measurements in the $r$-$\phi$ 
view \cite{SVX}. Surrounding the SVX is the COT, an open cell drift chamber 
with 96 layers of sense wires \cite{COT}. A sample rich in charm and beauty 
hadrons is selected by a three-level displaced track trigger. At level 1, 
tracks are reconstructed in the COT by the track trigger processor 
(XFT) \cite{XFT}. The trigger requires two tracks with transverse momenta 
$p_{T} > 2$ $\GeVc$ and the scalar sum $p_{T1}+ p_{T2} > 4.0$ $\GeVc$. 
The level 2 silicon vertex tracker (SVT) \cite{SVT} associates  SVX II 
$r$-$\phi$ position measurements with XFT tracks, providing a precise 
measurement of the track impact parameter ($d_{0}$), i.e. the distance of 
closest approach of the track helix to the beam axis in the transverse plane. 
Decays of heavy flavor particles are identified by requiring two tracks with 
$0.12 ~\mm < d_{0} < 1 ~\mm$ and an opening angle in the transverse plane 
$2^{\circ} < |\Delta \phi| < 90^{\circ}$. A requirement $L_{xy} > 0.2$ $\mm$ 
is also applied, where $L_{xy}$ is defined as the distance in the transverse 
plane from the beam line to the two-track vertex projected onto the two-track 
momentum vector. The level 3 trigger receives the complete detector 
information and performs a full event reconstruction applying selection 
similar to level 1 and level 2 on offline quality quantities.
\begin{figure}
\includegraphics[scale=0.4]{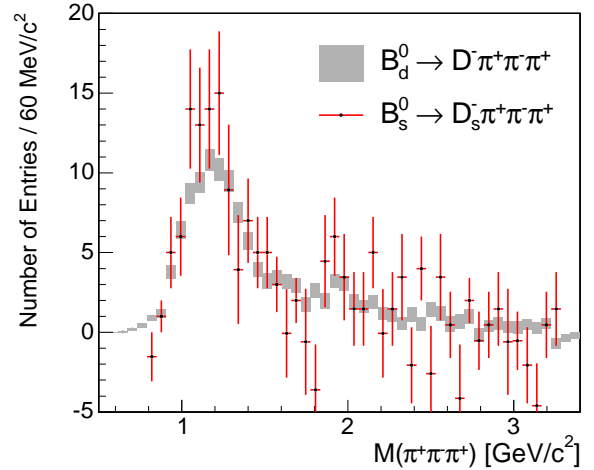}
\caption{ Comparison of the sideband subtracted mass spectrum of $\tripi$ 
from $\Bd$ and $\Bs$ decays. Only the $\Ds \to \phi \pi$ channel is used 
for $\Bs$. The $\Bd$ histogram is normalized to $\Bs$.}  
\label{fig:tripiMass}
\end{figure}
\begin{figure*}
\includegraphics[width = 0.65\textwidth, height =0.35\textheight]{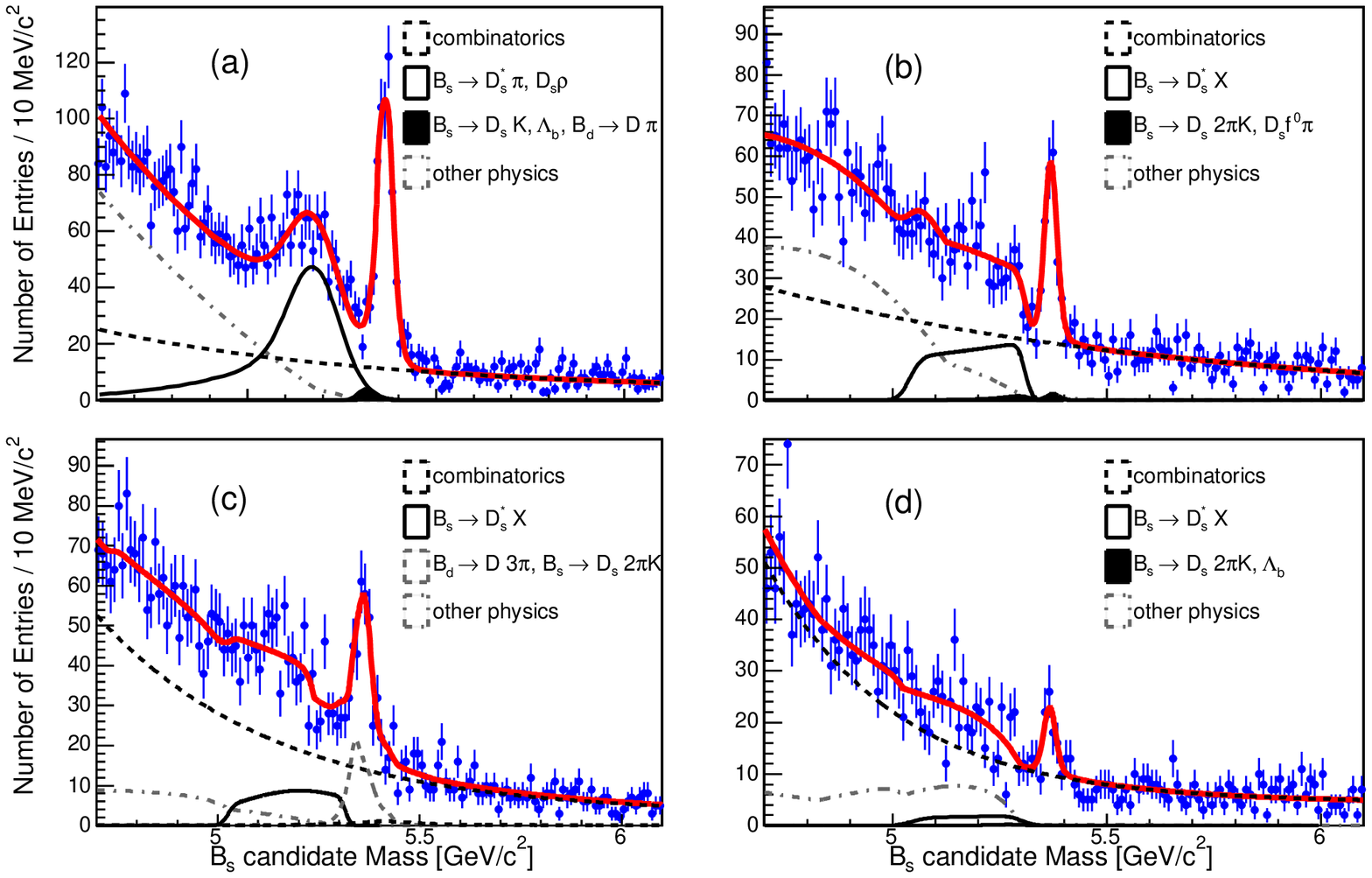}\nolinebreak
\includegraphics[width = 0.35 \textwidth, height = 0.35 \textheight]{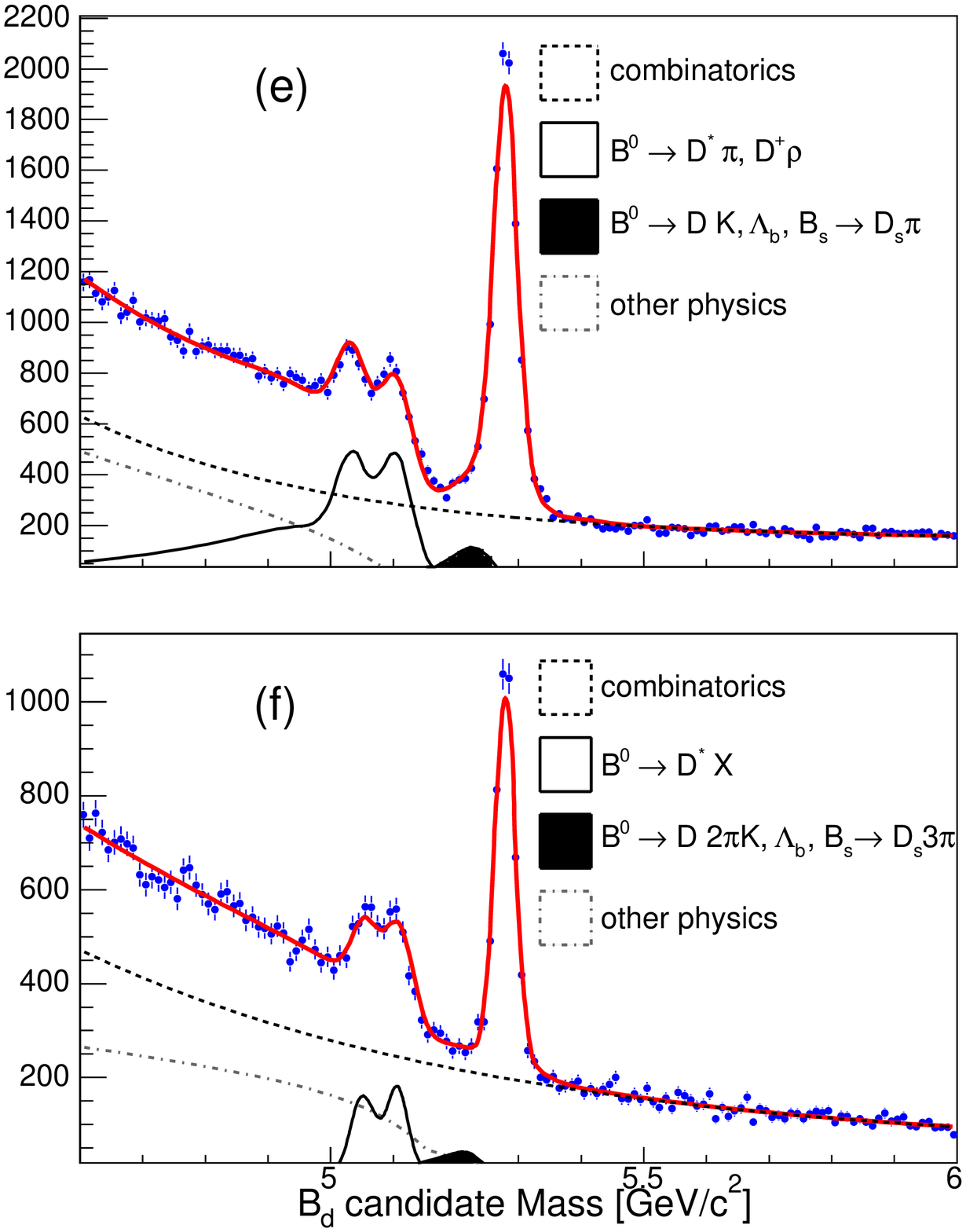} 
\caption{ 
Mass spectra for (a) $\Bs \to \Ds \pip$ (b) $\Bs \to \Ds(\phi\pim) \tripi$ 
(c) $\Bs \to \Ds(K^{*0}K^{-}) \tripi$ (d) $\Bs \to \Ds(\pim\pim\pip) \tripi$ 
(e) $\Bd \to \Dd \pip$ (f) $\Bd \to \Dd \tripi$. The ``other physics'' 
category corresponds to the inclusive $B \to \Ds X$ and $B \to \Dd X$ decays. 
}
\label{fig:massPeaks}
\end{figure*}

$B$ candidate reconstruction starts with a collection of tracks.
No particle identification is explicitly used, and tracks are assumed
 to be either a pion or kaon to match the reconstruction hypothesis. 
A set of unique track combinations making the  
$\phi, K^{*0}, \Dd, \Ds$, and $B$ candidates is formed. The track combinations 
reconstructed in 3 dimensions must be consistent with forming a vertex, 
and combinations that fall outside a wide mass window around the mass of 
the respective meson are rejected.  

The Monte Carlo simulation is an essential part of this analysis. It is used 
to optimize the selection cuts, model signal and background, and to 
study the trigger and reconstruction efficiency. We generate single $B$ 
hadrons with the program {\sc BGenerator}\cite{NDE}. The $B$-hadron decays 
are simulated with {\sc EvtGen} \cite{EVTGEN}. This package has been 
extensively tuned by experiments at the $\Upsilon(4S)$ resonance and 
reflects the measured properties of $B$ and $D$ meson decays. The simulation 
is validated by comparing kinematic quantities of the decays and quantities 
relevant for the trigger selection with data. 

The selection requirements used to reject combinatorial background are 
optimized by maximizing $\mathcal{S / \sqrt{S + B}}$ for each mode 
individually. The number of signal events ($\mathcal{S}$) is derived from 
a Monte Carlo simulation of the CDF II detector and trigger. The number of 
background events ($\mathcal{B}$) is estimated using
data in the high-mass sideband interval, $m(B)+10~\sigma(B)$ to 
$m(B)+16~\sigma(B)$, where $m(B)$ is the fitted mass and 
$\sigma(B) \approx 15~\MeVc^2$ is the width of the signal peak. This sideband 
represents the combinatoric background underneath the signal peak. Selection 
requirements include cuts on the impact parameter of the $B$ meson, 
the $\chi^2_{r-\phi}$ \cite{CHI} of the $B$ vertex fit in the transverse 
plane, the $p_{T}$ of the pion from the $B$ decay in $\aB \to \aD \pip$, 
and a minimum $p_{T}$ requirement of the tracks for decays with 6 tracks 
in the final state. We exploit the narrow $\phi \to K^{+}K^{-}$ resonance 
and $K^{*0} \to K^{+}\pim$ resonance to suppress background by requiring 
$1010$ $\MeVc^2 < m(\phi) < 1029$ $\MeVc^2$ and 
$840$ $\MeVc^2 < m(K^{*0}) < 940$ $\MeVc^2$. There are also requirements 
on $L_{xy}/\sigma(L_{xy})$ - the significance of the measurement of $L_{xy}$ 
for $B$ and $D$ vertices. 
\begin{table*}[ht]
\begin{center}
\begin{ruledtabular}
\begin{tabular}{lcccc}
Decay & Yield & $\epsilon(\Bs)/\epsilon(\Bd)$ & $\mathcal{B}(\Dd)/\mathcal{B}(\Ds)$ & $f_{s}/f_{d} \cdot BR(\Bs)/BR(\Bd)$ \\
\hline

$\Ds (\phi \pim)\pip$ &  $\nBsDsPiTotal \pm \eBsDsPiTotal$  & $\effRatiodspi \pm \effRatiodspiErr$ & $\dpdsphipir \pm \dpdsphipirErr $ & $\bsdspiRes \pm \bsdspiResStat (stat) \pm \bsdspiResSyst (syst)$ \\
$\Ds (\phi \pim)\tripi$ &  $\nBsDsTriPiTotal \pm \eBsDsTriPiTotal$  & $\effRatiodstripi \pm \effRatiodstripiErr$ & $\dpdsphipir \pm \dpdsphipirErr $ & $\bsdstripiRes \pm \bsdstripiResStat (stat) \pm \bsdstripiResSyst (syst)$   \\
$\Ds (K^{*}K^{-})\tripi$ &  $\nBsKSKTotal \pm \eBsKSKTotal$  & $\effRatiodsksk \pm \effRatiodskskErr$ & $\dpdskskr \pm \dpdskskrErr$ & $\bsdskskRes \pm \bsdskskResStat (stat) \pm \bsdskskResSyst (syst)$ \\
$\Ds (\pim\pip\pim)\tripi$ &  $\nBsTriPiTotal \pm \eBsTriPiTotal$  & $\effRatiodsTripi \pm \effRatiodsTripiErr$ & $\dpdstripir \pm \dpdstripirErr$ & $\bsdsTripiRes \pm \bsdsTripiResStat (stat) \pm \bsdsTripiResSyst (syst)$ \\

\end{tabular}
\end{ruledtabular}
\end{center}
\caption{ Summary of event yields, ratios of efficiencies, and individual 
branching ratio measurements. The uncertainties listed on the yield and 
the ratio of efficiencies are statistical only. 
$\mathcal{B}(\Dd)/\mathcal{B}(\Ds)$ is the ratio of 
$\mathcal{B}(\Dd \to K^{+}\pim\pim)$ to the corresponding branching fraction 
of the $\Ds$ meson ($\mathcal{B}(\Ds \to \phi \pim, \phi \to K^{-}K^{+})$, 
$\mathcal{B}(\Ds \to K^{*0}K^{-}, K^{*0} \to K^{-} \pip)$ or 
$\mathcal{B}(\Ds \to \pim \pip \pim)$). }
\label{tab:comboyields}
\end{table*} 

The assumptions on the relative contributions of resonant $a_{1}, \rho \pi$, 
and non-resonant $\tripi$ in the $\Bd \to \Dd \tripi$ signal affect 
$R(D 3\pi)$ because Monte Carlo simulation shows that their reconstruction 
efficiencies differ. We find that the contribution of the $a_{1}$ resonance 
is dominant. The $\tripi$ mass distributions were compared in data between 
$\Bd$ and $\Bs$ mesons and are compatible within statistics, as shown 
in Fig.\ref{fig:tripiMass}. The resonant fractions in $\Bs \to \Ds \tripi$
and $\Bd \to \Dd \tripi$ decays are assumed to be identical.  

To extract $R(D 3\pi)$ (or equivalently $R(D \pi)$), we follow 
the general procedure described in Ref. ~\cite{IVAN}, using the following
formula:

\begin{equation}\label{eq:rates}
  R(D 3\pi) = \frac{f_{d}}{f_{s}} \cdot \frac{\epsilon(\Bd)}{\epsilon(\Bs)}
    \cdot \frac{\mathcal{B}(\Dd)}{\mathcal{B}(\Ds )} \cdot \frac{N(\Bs)}{N(\Bd)},
\end{equation}
where $N(\Bs)$ and $N(\Bd)$ are the measured signal yields, 
$\epsilon(\Bd)/\epsilon(\Bs)$  is the ratio of trigger and reconstruction 
efficiencies extracted from Monte Carlo simulation, $f_{d}/f_{s}$ is the 
ratio of $b$ quark fragmentation fractions into $\Bd$ and $\Bs$ mesons, 
and $\mathcal{B}(\Dd)/\mathcal{B}(\Ds)$  is the ratio of the world average 
values for branching fractions of $\Dd$ and $\Ds$ mesons into the 
reconstructed final states \cite{PDG}.   

The yields, $N(\Bs)$ and $N(\Bd)$, are extracted from the mass spectra 
in Fig.\ref{fig:massPeaks} using a binned likelihood fit and are summarized in 
Table \ref{tab:comboyields}. In the normalization modes $\Bd \to \Dd \pip$ and 
$\Bd \to \Dd \tripi$ we observe $\nBdDPiTotal \pm \eBdDPiTotal \stat$ and 
$\nBdDTriPiTotal \pm \eBdDTriPiTotal \stat$ signal candidates respectively. 
The signal peaks are modeled with a sum of two Gaussians with the same mean 
values but different widths. We follow the treatment of backgrounds as done 
in Ref. \cite{IVAN}. The combinatorial background is modeled with a sum of 
an exponential function and a constant. The shapes of other physics 
backgrounds are modeled using Monte Carlo simulation, and their 
parameterization is fixed in the fits to the mass spectra. 

For $\Bd$ meson candidates, the structure below the main $B$ peak has two 
pronounced peaks. This structure is the result of a 
$\Bd \to D^{*-} \pip (\tripi)$ decay followed by a $D^{*-} \to \Dd \pi^{0}$ 
decay. Angular momentum conservation in $B$ decay implies a spin alignment 
of the $D^{*-}$ that, combined with the low $Q$ value of of its decay,
leads to the two distinct peaks  in the $\Dd \pip(\tripi)$ mass distribution 
seen in the data. In the case of the analogous $\Bs \to D^{*-}_{s} \tripi$ 
decay, the $D^{*-}_{s}$ decays to $\Ds$ and a photon 94$\%$ of the 
time \cite{PDG}. This produces a single, relatively broad structure as 
seen in our data. None of these backgrounds contribute under the signal peak.

There are several backgrounds whose mass distributions peak near the signal 
region and must be subtracted. They are the Cabibbo suppressed decays 
$\aB \to \aD \Kp$ and $\aB \to \aD \Kp \pim\pip$, and mis-reconstructed 
baryon decays $\Lambda_{b}^{0} \to \Lambda_{c}^{+} \pim$ and 
$\Lambda_{b}^{0} \to \Lambda_{c}^{+} \pim  \pip \pim$. The ratio of 
Cabibbo-suppressed $\aB \to \aD \Kp$ background to the corresponding signal 
was fixed to the world average ratio of branching fractions \cite{PDG}. 
The ratio of Cabibbo-suppressed $\aB \to \aD \Kp \pim\pip$ decay to the 
signal is fixed to $ \frac{|V_{us}|^2}{|V_{ud}|^2} \approx 0.05$. The 
fraction of $\Lambda_{b}$ background was fixed using the recent CDF 
measurement of $\mathcal{B}(\Lambda_{b}^{0} \to \Lambda_{c}^{+} \pim)/\mathcal{B}(\Bd \to \Dd \pip)$ \cite{LAMBDA}. In all the cases, the ratios are 
corrected for the relative trigger and reconstruction efficiencies. 

In the case of the $\Bs \to \Ds \tripi$ decay, there is a reflection from 
$\Bd \to \Dd \tripi$. If one of the pions from a $\Dd \to K^{+} \pim \pim$ 
decay is reconstructed 
as a kaon, a peak is produced under the $\Bs$ signal region. These events 
contribute in the $\Bs \to \Ds \tripi$, $\Ds \to K^{*0}K^{-}$ decay because 
the $K^{*0}$ resonance is broad. The fraction of $\Bd \to \Dd \tripi$ events 
under $\Bs \to \Ds \tripi$ peaks is calculated from the observed number of 
$\Bd$ mesons. The number of $\Bd \to \Dd \tripi$ events under the 
$\Bs \to \Ds (K^{*0}K^{-}) \tripi$ mass distribution is estimated to be 
141 $\pm 6 \stat$. The systematic uncertainty assigned due to the 
$\Bd \to \Dd \tripi$ background subtraction is dominant in this channel 
and is a part  of  the ``$\Bs$ fit model'' (see Table ~\ref{tab:bsSystList}). 
For $\Bs \to \Ds \tripi$ with $\Ds \to \phi \pim$ the corresponding $\Bd$ 
background fraction is very small due to the narrow width of the $\phi$. 
The contamination of the double charm $\Bd \to \Dd D^{+}_{s}$, with 
$D^{+}_{s} \to \pip \pim \pip$, is estimated to be $\approx$ 1\% of the 
$\Bd \to \Dd \tripi$ signal and is subtracted from the measured yield. 
The contamination of $\Bs \to \Ds D^{+}_{s}$ in the $\Bs$ signal is found 
to be negligible. Applying a cut on the mass of three pions in the 
$\Bs \to \Ds \tripi$ decay around the mass of the $\Ds$ meson does not 
change the measured yields.  

Systematic uncertainties due to the fitting procedure are estimated by 
comparing the fitted yields after changing the mass range in which the fit is 
performed and also by varying the parameters of functions describing the 
backgrounds. The systematic uncertainty on the ratio of efficiencies 
$\epsilon(\Bs)/\epsilon(\Bd)$ comes from various physics sources. In all 
cases, the systematics were estimated by observing the change in the ratio 
of efficiencies when the effect was considered. The effect of the choice of 
$B$ meson spectrum used by Monte Carlo simulation was determined by 
reweighting the Monte Carlo events with $p_{T}$ spectrum based on NLO 
calculations \cite{NDE} to match the $p_{T}$ spectrum measured at CDF 
\cite{BISHAI}. To estimate the systematic uncertainty due to $B$ and $D$ 
lifetimes we varied the assumed lifetime of $B$ and $D$ meson in signal 
Monte Carlo within world average values \cite{PDG}. The composition 
systematic applies to the decay $\Bs \to \Ds(\pim\pip\pim) \tripi$ only 
and is due to the limited knowledge of the resonances in 
$\Ds\to \pim \pip \pim$ decay. We assign a systematic uncertainty due to 
the unknown resonance structure of the $\tripi$ system in $\aB \to \aD \tripi$
 decay by varying the fraction of $a_{1}$ component in both $\Bd$ and $\Bs$ 
signal Monte Carlo in a range consistent with the observed shape. The 
systematic uncertainties are summarized in Table ~\ref{tab:bsSystList}.

The results of the measurements for $R(D 3\pi)$ are summarized in 
Table \ref{tab:comboyields}. To average the results of the measurements
by the expected yield we use Eq.(\ref{eq:rates}), where $N(\Bs)$ is a sum 
of the yields in three $\Bs$ channels and $\epsilon(\Bs)$ is a linear 
combination of Monte Carlo efficiencies multiplied by the ratio of the 
branching fraction of $\Ds$  decay in a given channel and 
$\mathcal{B}(\Ds \to \phi \pi)$. Using 
$f_{s}/f_{d} = \fsOverfd \pm \fsOverfdError$ \cite{PDG}, we obtain: 
\begin{equation}
	R(D 3\pi)
	= \bsdstripiComboResCorr \pm \bsdstripiComboStatCorr \stat 
		\pm \bsdstripiComboSystCorr \syst \pm 
	\bsdstripiComboBrCorr \br \pm \bsdstripiComboPrCorr \pr
\nonumber
\end{equation} 
The $\br$ and $\pr$ uncertainties refer to the uncertainty on the $D$ meson 
branching 
fractions and the ratio of fragmentation fractions $f_{s}/f_{d}$. 

	Using Eq.(\ref{eq:rates}) and the input from 
Table \ref{tab:comboyields}, we obtain:
\begin{equation}
 		R(D \pi)	
	  = \bsdspiResCorr \pm \bsdspiStatCorr \stat  \pm 
	\bsdspiSystCorr \syst \pm \bsdspiEBrCorr \br \pm 
	\bsdspiEPrCorr \pr \nonumber
\nonumber 
\end{equation}
The measurement of $R(D \pi)$ is consistent with the previous CDF 
measurement \cite{IVAN} and supersedes that result with a statistical 
uncertainty reduced by a factor of two. This measurement is 
consistent within uncertainties with the theoretical prediction of 
$\Colangelo \pm \ColangeloErr$ \cite{DIAGRAMS}. 

In conclusion, we have presented the first measurement of the ratio of 
branching fractions $R(D 3\pi)$. We also have measured the ratio of 
branching fractions $R(D \pi)$ improving the statistical uncertainty by 
more than a factor of two. Using the world 
average values for $\mathcal{B}(\Bd \to \Dd \pip)$ and 
$\mathcal{B}(\Bd \to \Dd \tripi)$ \cite{PDG} we find  
$\mathcal{B}(\Bs \to \Ds \pip) = [\brBsDsPi \pm \brBsDsPiStatErr \stat 
\pm \brBsDsPiSystErr \syst] \times 10^{-3}$ and $\mathcal{B}(\Bs \to 
\Ds \tripi) = [\brBsDsTriPi \pm \brBsDsTriPiStatErr \stat \pm 
\brBsDsTriPiSystErr \syst] \times 10^{-3}$. 
\begin{table}[ht]
\begin{center}
\begin{ruledtabular}
\begin{tabular}{lcc}
  &  \multicolumn{2}{c} {Syst. Uncertainty[\%]} \\ 
 Effect &  $B \to D \pi$ & $B \to D 3\pi$ \\
\hline

$B$ $p_T$ spectrum & $\pm 3.0$   & $\pm 3.0$ \\
$ \Bs $ lifetime &  $\pm 2.1$    & $\pm 2.1$ \\
$3\pi$ resonance structure & n/a & $\pm 2.5$ \\
$\Ds \to 3\pi$ composition & n/a & $\pm 3.0$ \\
trigger simulation & $\pm 1.2$  & $\pm 1.1$ \\
$\Bd$ fit model & $\pm 0.5$ & $\pm 1.3$ \\
$\Bs$ fit model & $ \pm 1.5$ & $\pm (3.6-9.3)$\\
\hline
Total & $\pm 4.2$ & $\pm (6.7-10.9)$ \\

\end{tabular}
\end{ruledtabular}
\end{center}
\caption{ Summary of the relative systematics uncertainties on $R(D\pi)$ 
and $R(D 3\pi)$. The range of values appearing in the second column reflects
the differences in the fit systematic of the three $\Bs$ decays.  
}
\label{tab:bsSystList}
\end{table}

We thank the Fermilab staff and the technical staffs of the participating 
institutions for their vital contributions. This work was supported by the 
U.S. Department of Energy and National Science Foundation; the Italian 
Istituto Nazionale di Fisica Nucleare; the Ministry of Education, Culture, 
Sports, Science and Technology of Japan; the Natural Sciences and Engineering 
Research Council of Canada; the National Science Council of the Republic of 
China; the Swiss National Science Foundation; the A.P. Sloan Foundation; 
the Bundesministerium f\"ur Bildung und Forschung, Germany; the Korean 
Science and Engineering Foundation and the Korean Research Foundation; 
the Particle Physics and Astronomy Research Council and the Royal Society, 
UK; the Russian Foundation for Basic Research; the Comisi\'on 
Interministerial de Ciencia y Tecnolog\'{\i}a, Spain; in part by the European 
Community's Human Potential Programme under contract 
HPRN-CT-2002-00292; and the Academy of Finland.



\begin{thebibliography}{99}
\bibitem{CONJ}
	Reference to the charge-conjugate decays is implied here 
and throughout the text.
\bibitem{REFSYSTEM}
     CDF II uses a cylindrical coordinate system in which $\phi$ is an 
azimuthal angle, $r$ is the radius from the symmetry axis, $y$ points up, 
and $z$ points in the proton beam direction with the origin at the center 
of the detector. The transverse plane is the plane perpendicular to the $z$ 
axis.
\bibitem{CHI}
	$\chi^{2}_{r-\phi}$ is a $\chi^{2}$-like goodness-of-fit quantity 
using only the track parameters in the transverse plane.
\end{thebibliography}
\end{document}